\documentclass[twocolumn,reprint,superscriptaddress, amsmath,amssymb, aps,pra]{revtex4-1}

\usepackage{graphicx,subfigure}
\usepackage{enumerate}
\usepackage{color}
\usepackage{dcolumn}
\usepackage{epstopdf}
\usepackage{multirow}
\usepackage{appendix}
\usepackage{braket}
\usepackage{cancel}
\usepackage[colorlinks, breaklinks, citecolor=blue, linkcolor=blue,  urlcolor={blue}]{hyperref}
\usepackage{breakurl}
\usepackage{amsfonts}
\usepackage{amsmath}
\usepackage{amssymb}
\usepackage{mathrsfs}
\usepackage{epsfig}
\usepackage{bm}

\usepackage{colortbl}
\definecolor{mygray}{gray}{.9}
\definecolor{darkblue}{rgb}{1,1,.70}
\definecolor{lightblue}{rgb}{1,1,.90}

\begin{document}

\title{Proposal for realizing and probing topological crystalline insulators in optical lattices }

\author{Jing-Xin Liu}
\email[Email: ]{602023220045@smail.nju.edu.cn}
\affiliation{National Laboratory of Solid State Microstructures and School of Physics, Nanjing University, Nanjing 210093, China}
\affiliation{Collaborative Innovation Center of Advanced Microstructures, Nanjing 210093, China}
\author{Jian-Te Wang}
\affiliation{National Laboratory of Solid State Microstructures and School of Physics, Nanjing University, Nanjing 210093, China}
\affiliation{Collaborative Innovation Center of Advanced Microstructures, Nanjing 210093, China}
\author{Shi-Liang Zhu}
\email[Email: ]{slzunju@163.com}
\affiliation{Key Laboratory of Atomic and Subatomic Structure and Quantum Control (Ministry of Education), Guangdong Basic Research Center of Excellence for Structure and Fundamental Interactions of Matter, School of Physics, South China Normal University, Guangzhou 510006, China}
\affiliation{Guangdong Provincial Key Laboratory of Quantum Engineering and Quantum Materials, Guangdong-Hong Kong Joint Laboratory of Quantum Matter, Frontier Research Institute for Physics, South China Normal University, Guangzhou 510006, China}
\affiliation {Quantum Science Center of Guangdong-Hong Kong-Macao Greater Bay Area, Shenzhen, China}

\begin{abstract}
   We develop a lattice model which exhibits topological transitions from $Z_2$ topological insulators to mirror symmetry-protected topological crystalline insulators by introducing additional spin-orbit coupling terms. The topological phase is characterized by the mirror winding number, defined within the mirror symmetry invariant subspace, which ensures the protection of gapless edge states and zero-energy corner states under specific boundary conditions. Additionally, we propose a feasible scheme using ultracold atoms confined in a stacked hexagonal optical lattice with Raman fields to realize the two-dimensional topological crystalline insulators. Detection of the mirror winding number in these systems can be achieved by implementing a simple quench sequence and observing the evolution of the time-of-flight patterns.
\end{abstract}

\maketitle


\preprint{APS/123-QED}

\section{Introduction}


Topological insulators (TIs) exhibit the bulk-boundary correspondence behavior and generate symmetry-protected edge states\cite{RevModPhys.82.3045,RevModPhys.83.1057}.The  bulk is insulating but supports gapless edge states, and these local boundary states are robust against perturbations as long as the corresponding symmetry is not broken.   Beyond the bulk-boundary correspondence, Benalcazar et. al. proposed the concept of electric multi-pole insulators\cite{doi:10.1126/science.aah6442,PhysRevB.96.245115}, which exhibit bulk-corner correspondence. In 2D, these insulators do not have edge states but instead have isolated corner states corresponding to quantized higher electric multi-pole moments. This extends the notion of correspondences between $d$-dimensional bulk and $(d-n)$-dimensional boundaries to the concept of higher-order TIs. Consequently, there is a new possibility that TIs, which were originally considered to be in the trivial phase, may actually hide richer higher-order phases\cite{PhysRevLett.118.076803,PhysRevLett.119.246401,PhysRevB.96.235424,PhysRevLett.120.026801,PhysRevB.97.205135,PhysRevLett.123.256402,PhysRevX.9.011012,PhysRevB.99.041301,PhysRevLett.123.216803,PhysRevLett.123.156801,PhysRevLett.124.136407,lee2020two,trifunovic2021higher}.

The appearance of corner modes is typically protected by crystalline symmetries or a combination of other symmetries. Therefore, we can use a broad class known as topological crystalline insulators (TCIs) to define systems with corner, edge, and hinge modes \cite{PhysRevLett.106.106802,hsieh2012topological}. Describing their topological properties involves creating topological classification tables that consider crystalline symmetry as well as all symmetries within the Altland-Zirnbauer tenfold classification scheme \cite{RevModPhys.88.035005,PhysRevB.88.075142,PhysRevLett.119.246401,PhysRevB.99.075105}. Furthermore, different mechanisms exist to distinguish and define the topological invariants of TCIs. One class can be straightforwardly classified by considering the bulk and boundary gap. However, for multi-pole higher-order topological insulators, the phase transition from a trivial insulator is characterized by a Wannier band gap\cite{PhysRevLett.125.037001,PhysRevLett.125.017001,PhysRevLett.125.166801,PhysRevX.10.041014,PhysRevB.102.121405,PhysRevB.102.155102,PhysRevB.102.100203,PhysRevResearch.3.013239,PhysRevResearch.2.033029,PhysRevB.98.045125}. For more detailed classification of higher-order topological insulator\cite{PhysRevResearch.5.L022032,PhysRevResearch.2.033029}, multi-pole moments \cite{doi:10.1126/science.aah6442,doi:10.1126/sciadv.aat0346,PhysRevLett.110.046404,PhysRevB.96.245115,PhysRevLett.119.246402,PhysRevB.95.235143} and bulk polarization \cite{PhysRevLett.118.076803} are used to identify Type-I higher-order TIs, while nested Wilson loops and Wannier band polarizations \cite{PhysRevResearch.3.013239} are applied to Type-II higher-order TIs. Therefore, the existence of topologically protected corner modes arises from various physical mechanisms. Experimental realizations of TCIs states have been widely reported in materials like bismuth \cite{schindler2018higher} and various synthetic systems such as phononic metamaterials\cite{serra2018observation,ni2020demonstration,PhysRevLett.128.224301,10.1063/5.0090596}, photonic metamaterials \cite{mittal2019photonic}, and electronic circuits \cite{imhof2018topolectrical,peterson2018quantized}.


Remarkably, ultracold atoms confined in optical lattices provide an excellent platform for emulating a wide range of systems in condensed matter physics \cite{APDWZhang2018,RevModPhys.80.885,lewenstein2007ultracold,bloch2012quantum,schafer2020tools}. Synthetic gauge fields and spin-orbit coupling (SOC) can be realized using various techniques, including trap rotation \cite{RevModPhys.81.647}, microrotation \cite{PhysRevLett.94.086803,PhysRevLett.100.130402,PhysRevA.82.013616,Mei2012}, and Raman laser-induced transitions \cite{RevModPhys.83.1523,lin2009synthetic,lin2011spin,SLZhu2011,PhysRevLett.102.130401,PhysRevLett.107.255301,PhysRevLett.95.010403,JZLi2022,jaksch2003creation,gerbier2010gauge,PhysRevLett.103.035301,LHHuang2016} By combining laser-induced tunneling and superlattice techniques, strong Abelian \cite{jaksch2003creation} and non-Abelian \cite{PhysRevLett.95.010403} gauge fields can be achieved, enabling the simulation of topological insulators and various topological matters  \cite{jotzu2014experimental,gerbier2010gauge,PhysRevLett.103.035301,PhysRevLett.97.240401,goldman2007quantum,PhysRevLett.101.246810,PhysRevLett.101.186807,PhysRevResearch.5.L012006,PhysRevA.79.053639,PhysRevLett.97.240401,XJLiu2007,Beeler2013,PhysRevLett.107.235301,LHHuang2016,PhysRevLett.107.145301,PhysRevLett.105.255302}. Especially, these technologies have been adopted for realizing atomic spin Hall effects~\cite{Beeler2013,PhysRevLett.97.240401,XJLiu2007}, quantum anomalous Hall effects~\cite{jotzu2014experimental,PhysRevLett.101.246810,PhysRevLett.101.186807} and 3D spin-orbit coupling\cite{doi:10.1126/science.aaf6689,PhysRevA.97.011605}. Based on this, there are proposals to realize second-order topological superconductors in cold atom systems\cite{PhysRevLett.123.060402,PhysRevA.107.043304,PhysRevLett.124.216601}.

In this article, we present an experimentally feasible scheme to achieve topological crystalline insulators with ultracold atoms in stacked hexagonal optical lattices possessing mirror symmetry. By incorporating additional gradient potential and Raman fields, we can establish a model in the quantum spin Hall phase with $s_x$ spin symmetry. Adjustment of nearest-neighbor  spin-flip hopping via Raman processes allows for the breaking of time-reversal $\mathcal{T}$ and $s_x$ symmetry while preserving mirror symmetry $\mathcal{M}_x$. In the TCIs phase, two $\mathcal{M}_x$ symmetry-protected corner states emerge, characterized by a $\mathbb{Z}_2$ class mirror winding number. This mirror winding number can be determined from the dynamics of time-of-flight (ToF) images~\cite{PhysRevLett.107.235301,PhysRevA.90.041601}, utilizing a straightforward quench sequence \cite{PhysRevLett.113.045303,doi:10.1126/science.aad4568,PhysRevResearch.5.L032016}.

The paper is structured as follows: In Section~\ref{sec:2}, we introduce a model that demonstrates non-trivial topology, edge states in a specific direction, and zero-energy corner states under certain boundary conditions. We explore the origin of corner states, the effective edge Hamiltonian, and the bulk topological invariant known as the mirror winding number. Section~\ref{sec:3} discusses the experimental setup involving optical lattices and Raman fields to achieve the mirror symmetry-protected TCIs phase. In Section~\ref{sec:4}, we outline a method to extract the mirror winding number from time-of-flight (ToF) images by manipulating the quench phase. Finally, Section~\ref{sec:5} concludes the paper with a summary of key findings and conclusions.

\section{Model}
\label{sec:2}

\begin{figure*}[tbph]
  \centering\includegraphics[width=14cm]{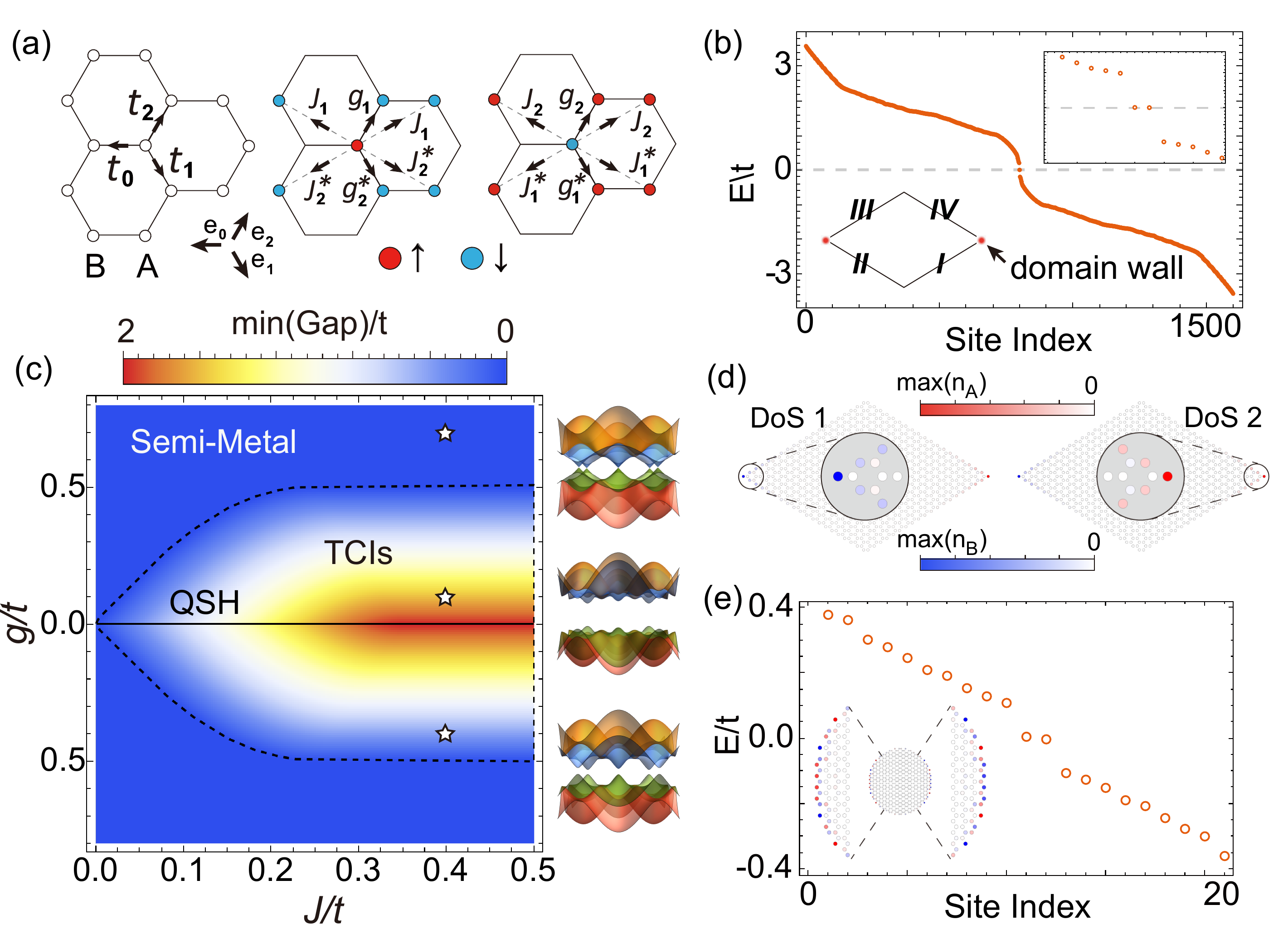}
  \caption{ (a) Hopping structure of the model. (b) Energy spectrum for a lattice with rhombus boundary.(c) Phase diagram: spin Hall phase $g=0$ (black solid line), topological crystalline insulator phase boundary (black dashed line). (d) Density distribution of zero-energy states (DoS). (e) Energy spectrum and DoS of a lattice with a circular boundary. }
  \label{fig:1}
\end{figure*}

    We consider  two-component fermionic atoms loaded in a tight-binding honeycomb lattice with spin-flip hopping terms, which can be viewed as a four-band model. The corresponding coupling terms are shown in Fig.\ref{fig:1}(a), and the tight-binding Hamiltonian can be expressed in three parts as follows:
\begin{widetext}
\begin{align}
H_{0} &= -\sum_{\mathbf{r}\in \mathbf{R}_A} \sum_{i} t_i c^\dag_{\mathbf{r}+\mathbf{e}_i} c_{\mathbf{r}} + \mathrm{h.c.}, \cr
H_{1} &= \sum_{\mathbf{r} \in \mathbf{R}_A} \left( c^\dag_{\mathbf{r}+\mathbf{e}_0 - \mathbf{e}_1} c_{\mathbf{r}} + c^\dag_{\mathbf{r}+\mathbf{e}_2 - \mathbf{e}_0} c_{\mathbf{r}} \right) (J_1 s^- + J_2 s^+) - \sum_{\mathbf{r} \in \mathbf{R}_B} \left( c^\dag_{\mathbf{r}+\mathbf{e}_0 - \mathbf{e}_1} c_{\mathbf{r}} + c^\dag_{\mathbf{r}+\mathbf{e}_2 - \mathbf{e}_0} c_{\mathbf{r}} \right) (J_1 s^- + J_2 s^+) + \mathrm{h.c.}, \\
H_{2} &= \sum_{\mathbf{r} \in \mathbf{R}_A} \left( g_1 c^\dag_{\mathbf{r} + \mathbf{e}_1} c_{\mathbf{r}} + g_2^\ast c^\dag_{\mathbf{r} + \mathbf{e}_2} c_{\mathbf{r}} \right) s^- + \sum_{\mathbf{r} \in \mathbf{R}_A} \left( g_2 c^\dag_{\mathbf{r} + \mathbf{e}_1} c_{\mathbf{r}} + g_1^\ast c^\dag_{\mathbf{r} + \mathbf{e}_2} c_{\mathbf{r}} \right) s^+ + \mathrm{h.c.}
\end{align}
\end{widetext}
where $c^\dag_{\mathbf{r}}$ is the creation operator for a fermion located at site $\mathbf{r}$, and $s^+ = (s^x + is^y)/2$ represents the spin raising operator. The unit direction $\mathbf{e}_j\ (j=0,1,2)$ is defined in Fig. 1a  . To create a model with non-trivial topology, we set $J_1 = J_2 = J_R + i J_I$ and $g_1 = -g_2 = i g$, where the parameters $J_{R,I}$ and $g$ describe the strengths of next-nearest neighbor and nearest neighbor  spin-flip hopping, respectively.

By performing a Fourier transform, the corresponding Bloch Hamiltonian is given by
\begin{equation}
H(\mathbf{k}) = \sum_{\mu,\nu=0}^3 \boldsymbol{d}_{\mu\nu}(\mathbf{k}) \cdot \left( \boldsymbol{s}_{\mu} \otimes \boldsymbol{\sigma}_{\nu}\right),
\end{equation}
where
\begin{equation}
\begin{split}
\boldsymbol{d}_{01} &= - \sum_{j} t_{j} \cos{(\mathbf{k} \cdot \mathbf{e}_j)},~~~\boldsymbol{d}_{02} = -\sum_{j} t_{j} \sin{(\mathbf{k} \cdot \mathbf{e}_j)}, \cr
\boldsymbol{d}_{13} &= 4 \cos{\frac{3 k_x}{2}} \left( J_R \cos {\frac{\sqrt{3} k_y}{2}} + J_I\sin {\frac{\sqrt{3} k_y}{2}} \right), \cr
\boldsymbol{d}_{21} &= 2 g \cos{\frac{k_x}{2}} \cos{\frac{\sqrt{3}k_y}{2}}, \cr
\boldsymbol{d}_{22} &= - 2 g \sin{\frac{k_x}{2}} \cos{\frac{\sqrt{3}k_y}{2}}.
\end{split}
\end{equation}

\subsection{Non-trivial bulk band topology with a mirror winding number}

In the simplest case where $g=0$ and $J_R= 0$, this model is in the quantum spin Hall phase with conserved $s_1$. By introducing the nearest-neighbor spin-flip terms, the phase diagram can be calculated from the bulk gap with eigenenergies $E = \pm\sqrt{A \pm 2\sqrt{B}}$ where $A = \boldsymbol{d}_{01}^2 + \boldsymbol{d}_{02}^2 + \boldsymbol{d}_{13}^2 + \boldsymbol{d}_{21}^2 + \boldsymbol{d}_{22}^2 $ and $B = (\boldsymbol{d}_{01}^2 + \boldsymbol{d}_{13}^2)\boldsymbol{d}_{21}^2 + (\boldsymbol{d}_{02}^2 + \boldsymbol{d}_{13}^2) \boldsymbol{d}_{22}^2 + 2 \boldsymbol{d}_{01} \boldsymbol{d}_{02} \boldsymbol{d}_{21} \boldsymbol{d}_{22}$. The gap results are shown in Fig.\ref{fig:1}(c). With the increase of $|g|$, in the TCIs region, we observe two isolated zero-energy corner states for diamond-shaped boundary in Fig.\ref{fig:1}(d). To understand this behavior, we first list all symmetries in this model. In this case, the model retains time-reversal symmetry $\mathcal{T} = s_3 \hat{K}$, particle-hole symmetry $\mathcal{C} = s_1 \sigma_3 \hat{K}$ and some crystalline symmetry, i.e., mirror symmetry $\mathcal{M}_{\mu} H(k_{\mu}) \mathcal{M}^{-1}_{\mu} = H(-k_{\mu})$, where $M_x = s_2 \sigma_1$ and $M_y = s_2$.

To demonstrate the unique topology arising from crystalline symmetry, we introduce a non-zero $J_R$ that breaks parity symmetry $\mathcal{P}$, time-reversal symmetry $\mathcal{T}$, particle-hole symmetry $\mathcal{C}$, and mirror symmetry $\mathcal{M}_y$. Only the combined symmetries $\mathcal{PT} = s_3 \sigma_1 \hat{K}$, $\mathcal{CP} = i s_1 \sigma_2 \hat{K}$, chiral symmetry $\mathcal{S} = s_2 \sigma_3$, and mirror symmetry $\mathcal{M}_x = s_2 \sigma_1$ are retained. According to the classification table of the $\mathcal{P}+\mathrm{AZ}$ class\cite{PhysRevB.96.155105}, this model falls into the $\mathrm{CI}$ class and possesses a $\mathbb{Z}_2$ topological invariant.

However, the presence of mirror symmetry $\mathcal{M}_x$ leads to symmetry-invariant subspaces, and the topological invariant can be described by the mirror-winding number (also known as the Zak phase). There exist two mirror-invariant lines, namely the $k_x = 0$ and $k_x = 2\pi$ planes. Along each line, we can define the subspace as a decomposition of $H(k)$, where the Hamiltonian can be represented as $H(0,k_y) = H^{+}(0,k_y) \oplus H^{-}(0,k_y)$. Here, $H^{\pm}(0,k_y)$ are spanned by the eigenstates ${\ket{\psi^{\pm}_n}}$ such that $\mathcal{M}_x \ket{\psi^{\pm}_n} = \pm \ket{\psi^{\pm}_n}$:
\begin{equation}
\begin{split}
H^{\pm}(0,k_y) &= \pm \boldsymbol{d}_{21} \sigma_0 + \boldsymbol{d}_{01} \sigma_1 \mp \boldsymbol{d}_{13} \sigma_2 \cr
\end{split}
\label{eq:1d}
\end{equation}
and this gives an effective 1D model. As $J_R$ increases, the bulk gap at $k_x=0$ will close and reopen. For each sector, this effective model is gapped when $|J_R| < \sqrt{3}|J_I|$. 

The presence of corner modes can be interpreted as resulting from a non-trivial winding in the mirror-invariant subspace, which can be quantified by the winding numbers\cite{PhysRevLett.124.166804}:
\begin{equation}
\nu_{\pm} = \frac{1}{2\pi}\int_{0}^{4\pi/\sqrt{3}} \mathrm{d}k_y~\mathrm{arg}\left[\boldsymbol{d}_{01}(k_y) \mp i \boldsymbol{d}_{13}(k_y) \right].
\end{equation}
Here it holds that $\nu_{+} + \nu_{-} = 0$ which relates to $\mathcal{PT}$ symmetry. For the case of half occupation, the mirror invariant can be defined as $\nu = (\nu_+ - \nu_-)/2$, where $\nu_+$ and $\nu_-$ are the winding numbers for two sectors. The wave functions of the two valence bands can then be expressed as
\begin{equation}
\ket{\psi_1} = \left[ 1,-e^{-i\theta}, -i e^{-i\theta},i \right]^T/2,
\label{eq:valence band}
\end{equation}
and  the wave function in the other sector is related as $\ket{\psi_2} = s_1 \ket{\psi_1}$.

We calculate the energy spectrum using two different open boundary conditions (OBC) and observed that the zig-zag case is always gapless, whereas the armchair case is gapped by a non-zero $g$. The system in the armchair boundary remains gapped in the topological crystalline insulator phase. However, for bulk semi-metal phases, the results are more complex. The OBC spectrum and edge gaps are presented in Appendix \ref{app:1}.

\subsection{Edge theory and corner modes}

Since edge states only exist along one direction, topologically protected corner states emerge at the intersection of two boundaries. To enhance clarity, we analyze our model using the edge theory \cite{PhysRevLett.99.196805,PhysRevLett.108.126807,PhysRevLett.121.096803,PhysRevB.98.045125,PhysRevB.105.L041105}, which is analogous to the Jackiw-Rebbi solitons \cite{PhysRevD.13.3398},  to provide a more intuitive understanding .

To illustrate the origin of corner states, we perform a Taylor expansion at the $\Gamma$ point for different boundaries. Considering that the model we are discussing involves a hexagonal lattice, we focus on the four boundaries of a rhombus, as depicted in Fig.\ref{fig:1}(d). As an example, by replacing $\sin{k_{\perp}} \to -i \partial_{r_{\perp}} $, $\cos{k_{\perp}} \to 1 - \frac{1}{2} \partial^2_{r_{\perp}} $, and $\sin{k_{\parallel}} \to k_{\parallel}$ while omitting insignificant $k_{\parallel}^2$ terms, the Hamiltonian at boundary I, shown in Fig.\ref{fig:1}(b) with $r^{(I)}_{\parallel} = (\sqrt{3}/2, 1/2)$ and $r^{(I)}_{\perp} = (- 1/2, \sqrt{3}/2)$, is given by
\begin{align}
H(-i\partial_{r_\perp},k_{\parallel}) &= H_0(-i\partial_{r_\perp},k_{\parallel}) + H_1(-i\partial_{r_\perp},k_{\parallel}), \\
H_0(-i\partial_{r_\perp},k_{\parallel}) &= \left( A_0 + A_2 \partial^2_{r_{\perp}} \right) \sigma_x + A_1 \partial_{r_{\perp}}  s_1 \sigma_3,
\end{align}
where $A_0 = -(t_0 + 2t_1)$, $A_1 = i 3 J_I/2$, and $A_2 = - (t_0 + 5 t_1)/8$. Here, we set $t_2 = t_1$. $H_1$ collects the remaining terms, and further details are provided in Appendix \ref{app:2}.

By solving the eigenvalue equation with zero energy, $H_0 \psi_{\alpha}(r_{\perp}) = 0$, under the boundary condition $\psi_{\alpha}(0) = \psi_{\alpha}(+\infty) = 0$, the zero-mode solution takes the form $\psi_{\alpha}(r_{\perp}) = e^{i k_{\parallel} r_{\parallel}} f(r_{\perp}) \chi_{\alpha}$, which satisfies the equation:
\begin{equation}
\begin{split}
 A_2 \partial^2_{r_{\perp}} f(r_{\perp}) + \alpha A_1 \partial_{r_{\perp}} f(r_{\perp}) + A_0 f(r_{\perp}) = 0.
\end{split}
\label{eq:1}
\end{equation}
Here we have introduced $s_x \sigma_z \chi_{\alpha} = \alpha \sigma_y \chi_{\alpha}$. The equation can be solved by the solutions $f(r_{\perp}) = \mathcal{N}_{r_{\perp}} \sin{(\kappa_1 r_{\perp})} e^{-\kappa_2 r_{\perp}}$ with $|\kappa_1| = \left| \sqrt{4A_0 A_2 - A_1^2}/2A_2 \right|$, $\kappa_2 = A_1/2A_2$ and normalization factor $\mathcal{N}_{r_{\perp}} = 2 \sqrt{\kappa_2\left( 1 + \kappa_2^2/\kappa_1^2 \right)}$. The condition $\alpha = +i$ corresponds to $\kappa_2 > 0$ and $f(+\infty) = 0$. Thus, the solution for the $\alpha = +i$ case represents the zero edge mode at boundary $\mathrm{I}$. The equations for the spinor part are given by $s_x \sigma_z \chi_{\alpha} = +i \sigma_x \chi_{\alpha}$, which yields
\begin{align}
\chi^{(\mathrm{I})}_1 = \ket{s_1 = +1}\ket{\sigma_1 = -1}, \\
\chi^{(\mathrm{I})}_2 = \ket{s_1 = -1}\ket{\sigma_1 = +1}.
\end{align}
Similarly, at the opposite boundary $\mathrm{III}$ with $\alpha = -i$, the spinor components are
\begin{align}
\chi^{(\mathrm{III})}_1 = \ket{s_1 = +1}\ket{\sigma_1 = +1}, \\
\chi^{(\mathrm{III})}_2 = \ket{s_1 = -1}\ket{\sigma_1 = -1}.
\end{align}

\begin{figure*}[tbph]
    \centering\includegraphics[width=16cm]{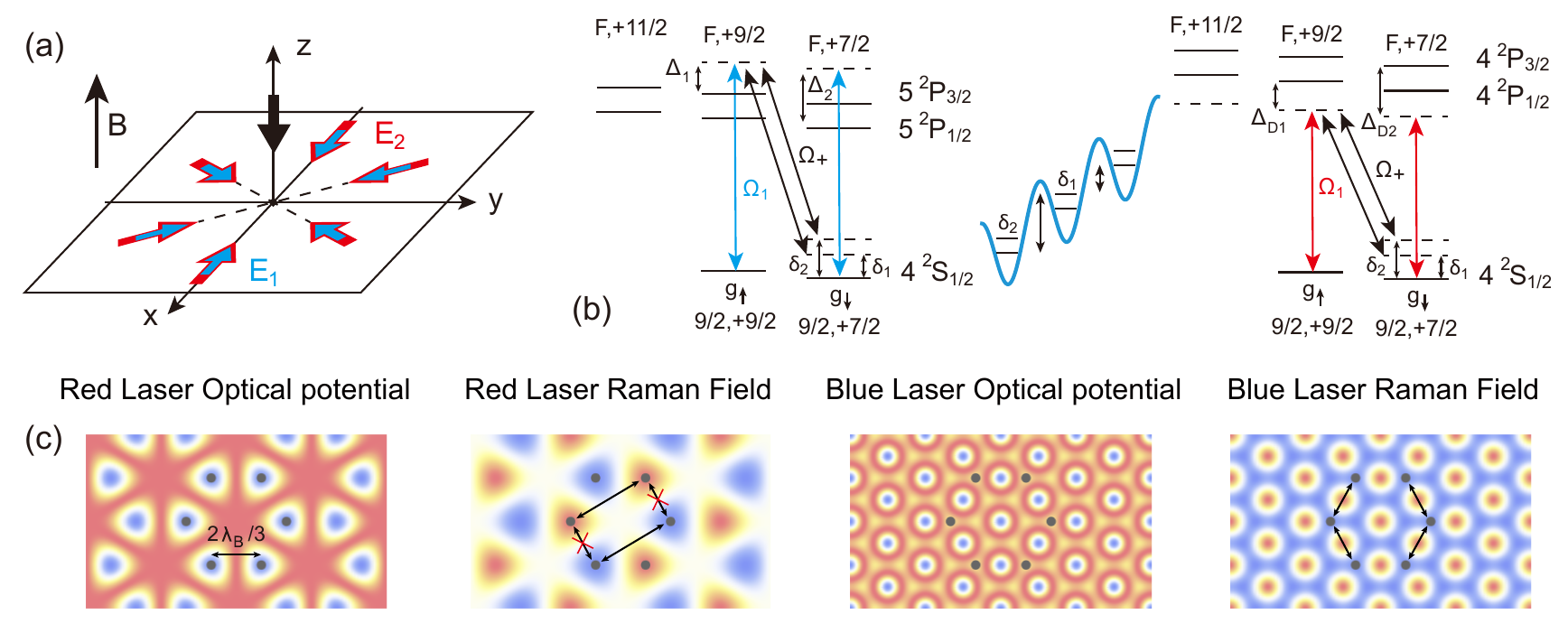}
    \caption{ (a) Experimental setup of a stacked optical lattice with blue and red laser fields. Additional laser fields propagate along the $\hat{z}$-direction for the Raman process. (b) Example of $~^{40}\mathrm{K}$ atom levels chosen to satisfy $\lambda_R/\lambda_B \approx 2$. A gradient potential generates a linear energy offset in the optical lattice along one specific direction. (c) Distribution of optical potential and Raman fields for blue and red lasers. }
    \label{fig:2}
\end{figure*}

The boundary Hamiltonian $H_{\mathrm{I}}$ with the zero boundary mode basis is given by
\begin{equation}
H_{\mathrm{I}} = v_z k_{\parallel} \tau_3 + m_x \tau_1.
\end{equation}
For the opposite boundary, we have $H_{\mathrm{III}} = -H_{\mathrm{I}} = - v_z k_{\parallel} \tau_3 - m_x \tau_1$. Similarly, we calculate the boundary Hamiltonians at boundaries $\mathrm{II}$ and $\mathrm{IV}$: $H_{\mathrm{II}} = -v_z k_{\parallel} \tau_3 + m_x \tau_1$ and $H_{\mathrm{IV}} = v_z k_{\parallel} \tau_3 - m_x \tau_1$.

The presence of $g$ introduces an effective mass term at each boundary. The difference in mass terms for the boundary massive Dirac fermion results in the Jackiw-Rebbi model \cite{PhysRevD.13.3398}, where the domain wall leads to corner states. This leads to corner states located at the intersection of boundaries $\mathrm{I},\mathrm{IV}$ and $\mathrm{II},\mathrm{III}$ as shown in Fig.\ref{fig:1}(b,d). 

Besides the above example of the rhombus boundary, we also depict the curved states with circular boundaries, and the results are shown in Fig.\ref{fig:1}(e). Due to the effective edge Hamiltonian having no mass term in specific direction, edge states preferentially form along the $\hat{y}$ direction. This explains the observed curved states  demonstrated  in Fig.\ref{fig:1}(e).

\section{Physical Realization}
\label{sec:3}

In this section, we present the details of deriving effective Hamiltonians based on realistic cold atom candidates and analyze the corresponding tight-binding model.

The formation of a hexagonal optical lattice \cite{PhysRevLett.101.246810,SLZhu2007,LMDuan2003,soltan2011multi,PhysRevA.80.043411} can be achieved through the intersection of three laser beams at a $120^\circ$ angle. We generate the hexagonal lattice and spin-orbit coupling (SOC) by stacking optical lattices with wavelengths $\lambda_R$ and $\lambda_B$, along with additional laser fields propagating along the $z$ direction as shown in Fig.\ref{fig:2}(a). To simultaneously match the spatial periods of the two optical lattices, the wavelengths of the red and blue lasers should satisfy $\lambda_R/\lambda_B = 2$. The total electric field of optical lattice is denoted as
\begin{equation}
\begin{split}
E_{OL}(\mathbf{r}) &= \sum_{n=1}^{2} E_n e^{-i \omega t/n} \sum_{j=1}^{3} \cos{\left( n \mathbf{k}_{j} \cdot \mathbf{r} + \frac{n\pi}{2}\right)} \hat{\mathbf{z}}, \cr
\end{split}
\end{equation}
where the wavevector $\mathbf{k}_1 = (1,0) k_0$, $\mathbf{k}_2 = (-\frac{1}{2},\frac{\sqrt{3}}{2}) k_0$, $\mathbf{k}_3 = (-\frac{1}{2},-\frac{\sqrt{3}}{2}) k_0 $ with the unit $k_0 = 2\pi/\lambda_B$.

The optical potential for atoms in the ground state can generally be expressed as $ V(\mathbf{r}) = u_s |\mathbf{E}|^2 + i u_{v} \left( \mathbf{E}^\ast \times \mathbf{E} \right) \cdot \mathbf{S}$, where the first term represents the scalar potential, and the second term denotes a vector light shift. The choice of laser wavelength can be addressed by selecting the $D_1$ line and $6^2S_{1/2} \to 7^2P_{3/2}$ transition of $^{133}\mathrm{Cs}$ atoms with a nuclear spin $I=7/2$, resulting in $\lambda_{D_1}^{(b)}/\lambda_1^{(b)} = 1.963 $, where $\lambda_1^{(b)}$ represents the wavelength of the $ 6^2S_{1/2}\to7^2P_{3/2}$ transition \cite{133Cs}. For the fermionic case, we can choose the $D_1$ line and $4^2S_{1/2} \to 5^2P_{3/2}$ transition of $^{40}\mathrm{K}$ atoms with a nuclear spin $I = 4$, giving $\lambda_{D_1}/\lambda_{1}^{(f)} = 1.904 $, where $\lambda_1^{(f)}$ represents the wavelength of the $4^2S_{1/2} \to 5^2P_{3/2}$ transition \cite{40K}.

Before discussing TCIs phases, we should first realize four-band topological insulators (TIs). The realization of a $\mathcal{T}$-symmetric model has been achieved in an optical lattice through the incorporating of a gradient magnetic field and Raman laser fields \cite{PhysRevLett.111.225301,PhysRevLett.111.185301,PhysRevLett.105.255302,DWZhang2016}. An approach we can consider involves introducing an additional gradient potential. This potential can facilitate the independent coupling of spins at different sites. By applying an artificial
electric field via accelerated motion of the optical lattice, we can derive the 2D effective Hamiltonian as follows:
\begin{equation}
H_{\mathrm{eff}}(\mathbf{r}) = \frac{\mathbf{p}^2}{2m} + V_{OL}(\mathbf{r}) + \mathcal{F} \cdot \mathbf{r} + \frac{\mu_B g_F}{\hbar}\mathbf{B}\cdot \mathbf{F}.
\label{eq: continue model}
\end{equation}

By adding laser fields $\mathbf{E}_{R1}$ and $\mathbf{E}_{R2}$ propagating along the $\hat{z}$ direction, the electric fields in the $z = 0$ plane are:
\begin{align}
\mathbf{E}_{R1} &= E_{R1} \left( e^{-i(\omega_{R1} + \delta_1) t } + e^{-i(\omega_{R1} + \delta_2) t } \right) (\hat{\mathbf{x}} + i \hat{\mathbf{y}}), \\
\mathbf{E}_{R2} &= E_{R2} \left( e^{-i(\omega_{R2} + \delta_1) t } + e^{-i(\omega_{R2} + \delta_2) t} \right) (\hat{\mathbf{x}} + i \hat{\mathbf{y}}),
\end{align}
where $\delta_1$ and $\delta_2$ compensate for the detuning caused by the gradient potential during different spin-flip processes. These laser fields can drive the $\sigma_+$ and $\pi$ transitions from the ground states $\ket{g_{\uparrow}} = \ket{9/2,+9/2}$ and $\ket{g_{\downarrow}} = \ket{9/2,+7/2}$ to all possible excited states that satisfy the selection rules. For the $D_1$ and $D_2$ lines in $^{40}\mathrm{K}$ atoms, the state $\ket{9/2,+7/2}$ couples to the excited states $\ket{F,+9/2}$ ($\sigma_+$). The red (blue) Raman fields, as depicted in Fig.\ref{fig:2}(b), correspond to the strengths $E_1$ ($E_2$) and are derived from the $D_1$ line ($4^2S_{1/2} \to 5^2P_{1/2}$) and $D_2$ line ($4^2S_{1/2} \to 5^2P_{3/2}$) transitions.

The strengths of the blue and red lasers' optical potentials, $V^{(1,2)}(\mathbf{r})$, and the corresponding Raman fields, $M_{R1,R2}(\mathbf{r})$, are illustrated in Fig.\ref{fig:2}(c). Details explanations of the optical lattice potential, the construction of Raman processes, and the relevant parameters can be found in Appendix \ref{app:3}. The overall optical potential is a result of contributions from both the blue and red lasers. Fig.\ref{fig:2}(c) depicts the potential distributions of the two lasers. By configuring the system such that $E_1 \gg E_2$, the blue laser's contribution to the potential becomes negligible and we don't need to consider the third sublattice degree. This setup ensures that the energy minimum points are located at the positions marked by black solid points, with a lattice constant of $2\lambda_B/3$.

To obtain the effective tight-binding model, we need to calculate the overlap integral of two Wannier-Stark states. First, let's consider a rough analysis: suppose there is no gradient potential. In such a honeycomb lattice, the Wannier function maintains $C_3$ symmetry (since the energy minima regions exhibit $C_3$ symmetry). In this case, the center of the Wannier function will coincide with the energy minimum point of the potential, as shown by the black point in Fig.\ref{fig:2}(c). The corresponding strengths of next-nearest-neighbor and nearest-neighbor spin-flip couplings can be calculated by overlapping two Wannier functions, taking into account the Raman field background. For the contributions of the red lasers, the nearest-neighbor spin-flip terms can be eliminated due to the $C_6$ anti-symmetry and $C_3$ symmetry of the Wannier function in Fig.\ref{fig:2}(c). By controlling the strength of $M_{R1}$, the next-nearest-neighbor spin-flip term can become sufficient large. The nearest-neighbor spin-flip hopping is controlled by the blue Raman field $M_{R2}$. The overlap of the next-nearest-neighbor Wannier function is small enough that we can disregard the next-nearest-neighbor terms contributed by the blue Raman field. The phase and strength of the Raman fields $M_{R1}$ and $M_{R2}$ can also be controlled independently by adjusting $E_{R1}$ and $E_{R2}$. By tuning the Rabi frequencies of the laser fields to satisfy $J_1 = J_2$ and $g_1 = -g_2 = -i g$, we can realize the effective lattice model described in Section \ref{sec:2}.

However, the presence of a gradient potential in Eq. \ref{eq: continue model} breaks the original $C_6$ symmetry of the potential while retaining mirror symmetry $\mathcal{M}_x$. Suppose the gradient potential along the $\hat{y}$ direction is not large enough, so the centers of the Wannier-Stark functions are only slightly displaced. In this situation, the overlap of two next-nearest-neighbor Wannier-Stark functions also results in opposite values for different sublattices. Additionally, there is an extra contribution from the nearest-neighbor spin-flip term caused by the blue Raman field. This introduces staggered coupling, which is also influenced by the $C_6$ anti-symmetry of the blue Raman field (as detailed in Appendix \ref{app:4}). The effect of these terms exhibits a similar form to the contribution from the red Raman field and will not break mirror symmetry $\mathcal{M}_x$.

\section{Probe Mirror winding number}
\label{sec:4}

In the $\mathcal{M}_x$-invariant subspace, the system can be viewed as two isolated 1D models, and its topology can be simply described by a single parameter $\theta(k_y)$ mentioned in Sec \ref{sec:2}. In the cold atom platform, this parameter of Bloch states can be extracted from interference patterns of ToF images\cite{PhysRevLett.113.045303,doi:10.1126/science.aad4568,APDWZhang2018,PhysRevLett.113.045303,doi:10.1126/science.aad4568}. In this section, we develop a method to probe the mirror winding number.

For this four-band model, the overall density distribution of the ToF image in momentum space can be calculated using the following expression:
\begin{equation}
\begin{split}
n(\mathbf{k}) &= f(\mathbf{k})\sum_{m,n} e^{-i \mathbf{k} \cdot (\mathbf{r}_{m} - \mathbf{r}_{n})} e^{i(\mu_{m,\sigma} - \mu_{n,\sigma})t} \cr
& \times \bra{G} C_m^\dag C_n \ket{G}, \cr
\end{split}
\label{eq: 4-TOF}
\end{equation}
where $C_n = [a_{n,\uparrow},b_{n,\uparrow},a_{n,\downarrow},b_{n,\downarrow}]^T$ and $a_{n,\sigma}$($b_{n,\sigma}$) represents annihilation operators for site $\mathbf{r}_n$ in sublattice A(B) with spin $\sigma$. The factor $f(\mathbf{k})$ denotes a broad envelope function determined by the momentum distribution of the Wannier function, and $\mu_{n,\sigma}$ indicates the global trap potential strength at site $n$. In the absence of a global trapping difference, we have $\mu_{n,\sigma} = \mu_s$. The state $\ket{G}$ represents the many-particle states with Fermi energy $E_f$ in an ideal periodic system without trapping differences. The ToF image can be expressed as $n(\mathbf{k}) = f(\mathbf{k})\bra{G}s_0\sigma_0 + s_0\sigma_1 + s_1\sigma_0 + s_1\sigma_1\ket{G}$, which contains all contributions from interference. In the case of periodic boundary condition (PBC), we sum the contribution of the two valence bands using Eq.(\ref{eq:valence band}). The contributions from the $s_1 \sigma_0$ and $s_1 \sigma_1$ terms are zero, yielding $n(k_y) = 2(1 - \cos{\theta(k_y)})$ theoretically. In this scenario, a quench involving the Pauli matrix $\sigma_3$ generates the evolution operator $\exp{(-i \sigma_3 \varphi/2)}$, which can extract $\theta(k_y)$ independently. This quench sequence leads to the ToF image into
\begin{equation}
n(0,k_y,\varphi) = 2(1 - \cos{\varphi} \cos{\theta(k_y)}).
\label{eq:theoTOF}
\end{equation}
and the parameter $\theta(k_y)$ can be extracted via adjusting the quench phases.

However, in a finite-size system, the ground states differ from Bloch waves in PBC. Consequently, the numerical results, including contributions from corner states, don't precisely match Eq. \ref{eq:theoTOF}. The actual results can be expressed similarly but with an additional background term due to OBC: $n_1(k_y,\varphi) = f(k)(1 - \cos{\varphi}\cos{\theta}) + b(k)$ where $b(k)$ is background caused by OBC. To account for the broadening caused by Wannier functions $f(k)$ and the background contributions from corner modes, we introduce another quench term $\exp{(-i \sigma_2 \xi/2)}$, where $n_2(k_y,\xi) = f(k)(1 - \cos{(\theta - \xi)}) + b(k)$. Consequently, we have $f(k) = (\mathrm{max}\{n_2(k)\} - \mathrm{min}\{n_2(k)\})$ and $b(k) = \mathrm{min}\{n_2(k)\}$.

\begin{figure}[tbph]
    \centering\includegraphics[width=8cm]{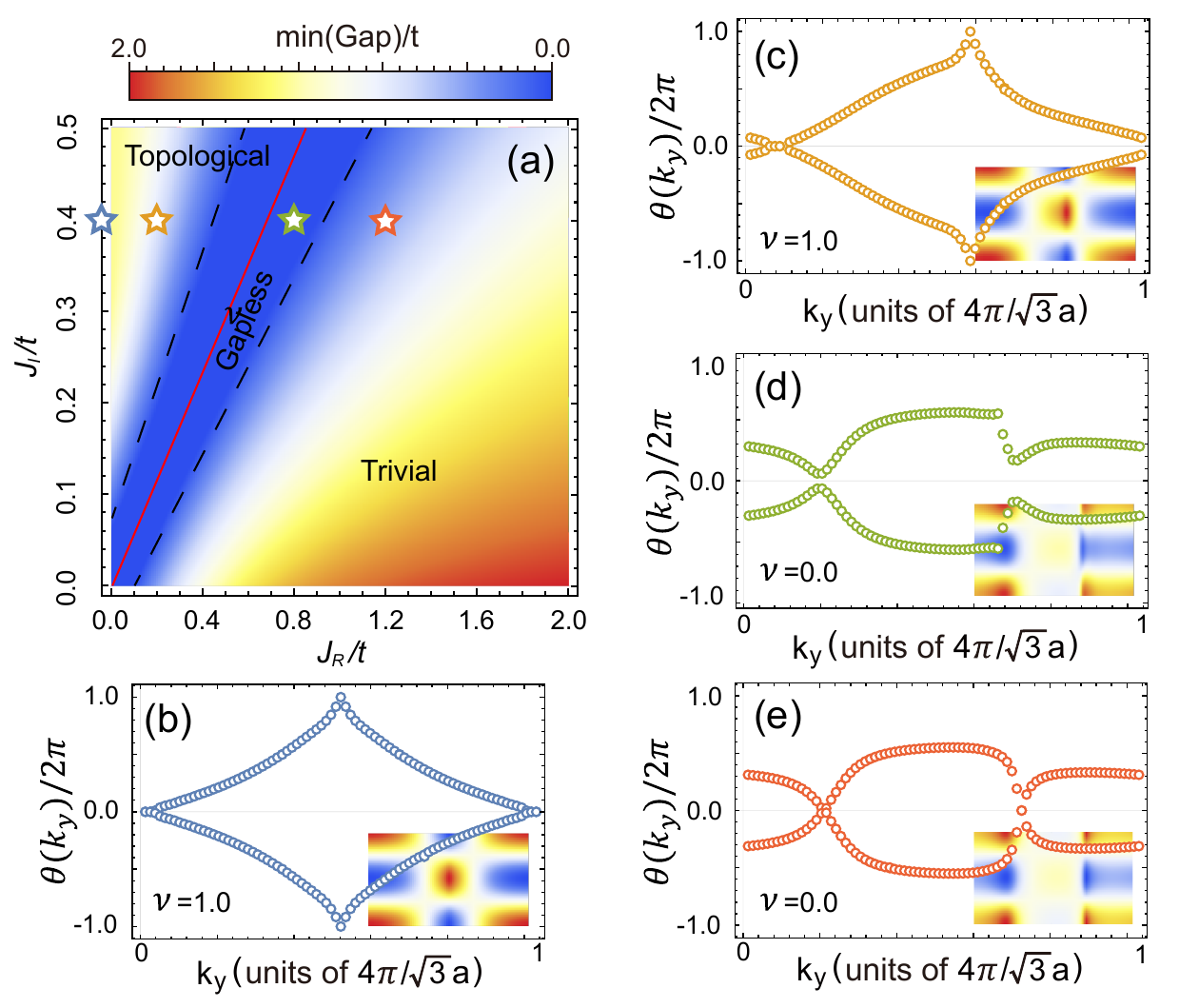}
    \caption{ (a) Bulk gap versus $J_R$ and $J_I$. The red solid line indicates the boundary of topological phase transition. The black dashed line marks the boundary of the gapless region. (b-e) Extracted $\theta_{\pm}$ for two $\mathcal{M}_x$-invariant subspaces. (b) Topological phase with $J/t = 0.4i$; (c) $J/t = 0.2 + 0.4i$; (d) gapless region with $J/t = 0.8 + 0.4i$; (e) trivial phase with $J/t = 1.2 + 0.4i$ in open boundary condition lattice with $8\times 8\times 2$ sites and $g/t = 0.2$. Subgraphs show modified ToF image evolution for different momentum $k_y$ by varying quench phases.}
    \label{fig:3}
\end{figure}

In Eq. \ref{eq:1d}, we see that the mirror winding number only relates to the ratio $|J_R/J_I|$. However, the additional spin-orbit coupling $g$ changes the bulk band structure from a gapped phase to gapless one. This implies that in some regions, we cannot extract the topological invariant because the valance bands are not well-defined. Therefore, even though the mirror winding numer remains non-trivial in some regions, we cannot extract it. We give the bulk gap for a system with constant nearest neighbour coupling $g$ in Fig.\ref{fig:3}(a). The red line $|J_R/J_I| = \sqrt{3}$ marks the boundary between topological and trivial phases. For regions with finite gap, we can extract information related to the topology of the valence bands.

From the ToF image, we can extract $\theta(k_y)$ as shown in Fig.\ref{fig:3}(b-e). In a system in the TCIs phase with a non-trivial mirror winding number and a finite bulk gap, $\cos{\theta(k_y)}$ falls within the range of $[-1,1]$. In other gapless regiona, the ToF image contains contributions from both valence and conduction bands, and the winding number is not well-defined. For a system in the trivial phase with a finite bulk gap, $\cos{\theta}$ cannot cover the entire value range and the mirror winding number must be zero.

\section{Conclusion}
\label{sec:5}

We have developed a lattice model that exhibits $Z_2$ TIs. By introducing additional SOC terms, we observe a phase transition from $Z_2$ TIs to mirror symmetry-protected TCIs. The topological invariant is defined in the mirror-symmetry-invariant subspace as the $\mathbb{Z}_2$ mirror winding number. This mirror symmetry protects gapless edge states at specific boundaries, and zero-energy corner states can be observed in particular open boundary conditions. We have proposed a feasible scheme using cold atoms with optical lattices and Raman fields to realize the TCIs, where all parameters in the Hamiltonian can be independently adjusted. The corresponding mirror winding number can also be probed by measuring the ToF image. This work provides a method to control and simulate various topological states of matter in cold atom systems.

\section*{Acknowledgment}

We thank Peng He and Y.Q. Zhu for helpful discussions. This work was supported by National Key R\&D Program of China (Grant No. 2022YFA1405300) and the National Natural Science Foundation of China (Grant No. 12074180).

\appendix

\section{Edge gap}
\label{app:1}

\begin{figure}[tbph]
    \centering\includegraphics[width=8cm]{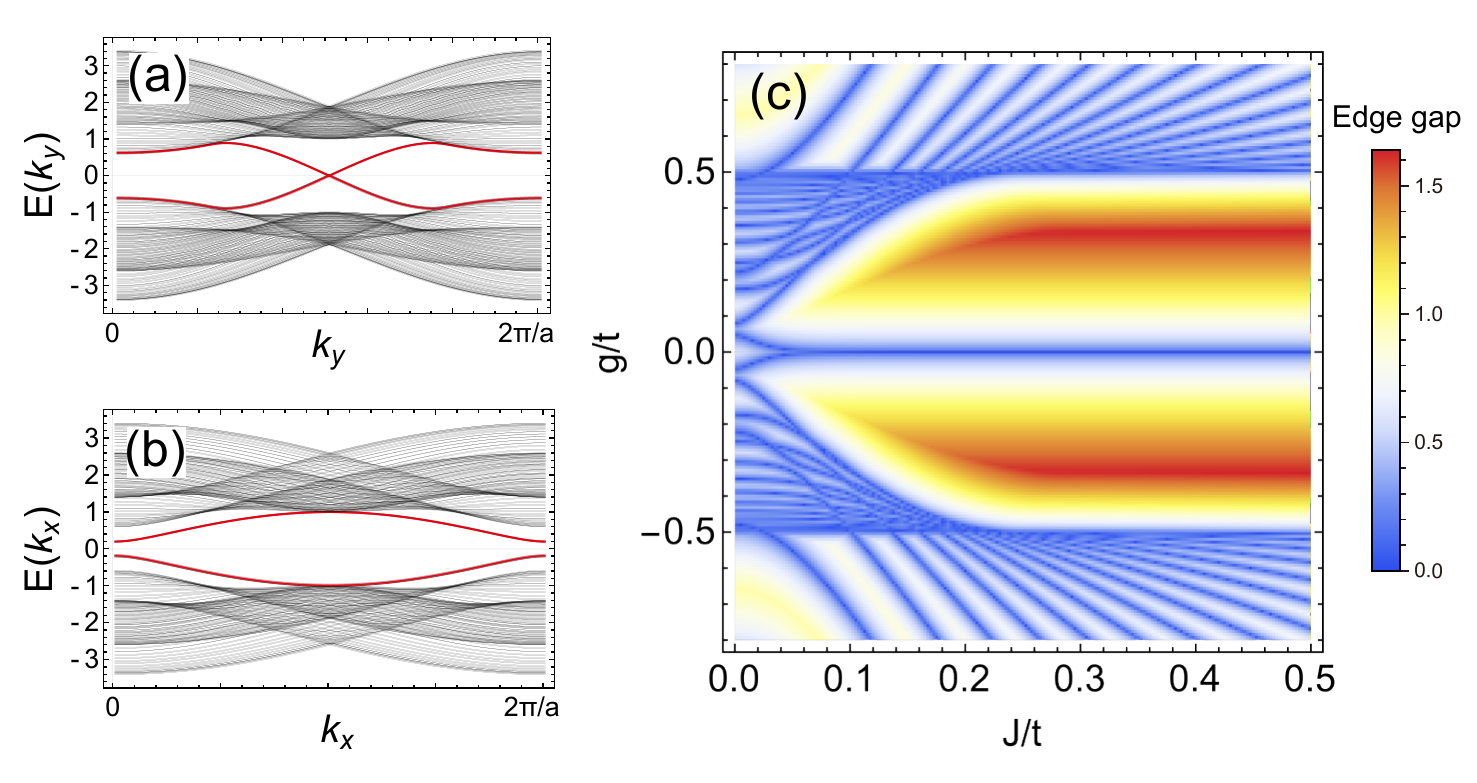}
    \caption{ Edge energy spectrum with zig-zag boundary in the x-direction OBC (a) and armchair boundary of y-direction OBC (b) with parameters $g/t = 0.2, J_I/t=0.4 $. (c) The edge gap of armchair boundary exhibits the same behavior as the bulk gap in the TCIs phase region. }
    \label{fig:4}
\end{figure}

Here, we present the edge band spectrum for zig-zag and armchair open boundary cases. With the introduction $g$, a gap opens at specific directions. From the edge gap diagram, we can clearly observe the vanishing of edge mode regions, which exhibits similar behaviors to the bulk gap diagram.

\section{Boundary Hamiltonian}
\label{app:2}

To demonstrate the origin of corner modes, we perform a Taylor expansion at the $\Gamma$ point for different boundaries. Given that the model we are discussing involves a hexagonal lattice, we can consider the boundaries with wave vectors $k_{1} = k_x,~k_{2} = k_x/2-\sqrt{3}k_y/2$ and $k_{3} = k_x/2+\sqrt{3}k_y/2$. We replace $\sin{k_{\perp}}$ with $\to -i \partial_{r_{\perp}}$ and $\cos{k_{\perp}}$ with $1$,  while omitting insignificant $k_{\parallel}^2$ terms.

The Hamiltonian at boundary I, with $r^{(I)}_{\parallel} = \sqrt{3}\hat{x}/2 + \hat{y}/2$ and $r^{(I)}_{\perp} = - \hat{x}/2 + \sqrt{3}\hat{y}/2$, can be decomposed into three parts:
\begin{equation}
\begin{split}
H_0 &= - \left[ t_0 + 2 t_1 + \frac{1}{8} (t_0 + 5t_1) \partial^2_{r_{\perp}} \right] \sigma_1 + \left( i \frac{3}{2} J_I \partial_{r_{\perp}} \right) s_1 \sigma_3 \cr
H_1 &= i \frac{\sqrt{3}}{4} k_{\parallel} (t_0 - t_1) \partial_{r_{\perp}} \sigma_1 + J_R \left( 2 + \frac{9}{8} \partial^2_{r_{\perp}} \right) s_1 \sigma_3 \cr\cr
& - \left[ \frac{\sqrt{3}}{2} J_I \left( 1 - \frac{9}{8} \partial^2_{r_{\perp}} \right) k_{\parallel} + i \frac{3\sqrt{3}}{4} J_R k_{\parallel} \partial_{r_{\perp}} \right] s_1 \sigma_3 \cr
& - \frac{1}{16} g \left[ \sqrt{3}\left( 8 + \partial^2_{r_{\perp}} \right)k_{\parallel} + 8 i \partial_{r_{\perp}} \right] s_2 \sigma_2 \cr
H_2 &= \frac{1}{16} (t_0 - t_1) \left[ \sqrt{3} \left( \partial^2_{r_{\perp}} + 8 \right) k_{\parallel} + 8i \partial_{r_{\perp}} \right] \sigma_2 \cr
& + \frac{1}{8} g \left[ 5 \partial^2_{r_{\perp}} + 2\sqrt{3} i \partial_{r_{\perp}} k_{\parallel} + 16 \right] s_2 \sigma_1 \cr
\end{split}
\end{equation}
where the first part $H_0$ has a zero-energy solution. The second and third parts, $H_1$ and $H_2$, include all other terms in a complex form. The zero-energy solutions are given in Sec.\ref{sec:2}. Here, we express the Hamiltonian in the subspace consisting of two zero-energy edges states:
\begin{equation}
H_{\mathrm{edge}} = \left( \begin{matrix}
\bra{\psi_1} H \ket{\psi_1} & \bra{\psi_1} H \ket{\psi_2} \\
\bra{\psi_2} H \ket{\psi_1} & \bra{\psi_2} H \ket{\psi_2} \\
\end{matrix}\right)
\end{equation}
At boundary I, only the terms involving $\sigma_2$ and $s_2 \sigma_1$ have non-zero values. Consequently, the contributions from $H_0$ and $H_1$ are zero in this subspace. In this scenario, the effective continuum Hamiltonian at the boundary can be represented as:
\begin{equation}
\begin{split}
H_{I} &= v_z k_{\parallel} \sigma_z + m_x \sigma_x.
\end{split}
\end{equation}
and this is essentially a massive Weyl fermion. The velocity and mass term can be calculated using these integrations:
\begin{equation}
\begin{split}
\int_{0}^{+\infty} \mathrm{d}r_{\perp} f^\ast(r_{\perp}) \partial_{r_{\perp}} f(r_{\perp}) &= 0 \cr
\int_{0}^{+\infty} \mathrm{d}r_{\perp} f^\ast(r_{\perp}) \partial^2_{r_{\perp}} f(r_{\perp}) &= - \mathcal{N}_y^2 \frac{\kappa_1^2}{4\kappa_2} \cr
\end{split}
\end{equation}

\section{Optical lattice and Raman field setting}
\label{app:3}

The red Raman fields, related to strength $E_1$, are contributed by the $D_1$ line and $D_2$ line transitions, where $E_{+1}^{(1)}(\mathbf{r}) = E_{R1}$ and $E_{0}^{(1)}(\mathbf{r}) = E_1 \sum_{j=1}^{3} \cos{ 2\mathbf{k}_{j} \cdot \mathbf{r}}$. The optical potential formed by red lasers in Fig.\ref{fig:2}(c) for spin $\tau$ is
\begin{equation}
\begin{split}
V^{(1)}_{\tau}(\mathbf{r}) &= \sum_{n=1}^{2} \sum_{F} \left( \frac{|\Omega^{D_n}_{\tau F,+1}|^2 }{\Delta_{D_n}} + \frac{|\Omega^{D_n}_{\tau F,0}|^2}{\Delta_{D_n}} \right). \cr
\end{split}
\end{equation}
where $\Omega_{\tau F}$ represents the Rabi frequency of all possible transitions from the state $\ket{4^2S_{1/2},9/2,+7/2}$. The Raman field caused in Fig.\ref{fig:2}(c) is
\begin{equation}
\begin{split}
M_{R1}(\mathbf{r}) &= \sum_F \left( \frac{\Omega^{D_2 \ast}_{\uparrow F,+1} \Omega^{D_2}_{\uparrow F,0}}{\Delta_{D_2}} + \frac{\Omega^{D_1 \ast}_{\uparrow F,+1} \Omega^{D_1}_{\uparrow F,0}}{\Delta_{D_1}} \right) \cr
& \left( \sum_{j=1}^{2} \delta( \Delta + \omega - \omega_{R1} - \delta_j(\mathbf{r})) \right) \ket{g_{\downarrow}}\bra{g_{\uparrow}} + h.c.\cr
\end{split}
\end{equation}
where $\Delta$ is the level splitting caused by the external Zeeman field and $\delta$ is the delta function. For example, the Rabi frequencies of the $D_1$ line mentioned in main text are given by $|\Omega_{\uparrow F,\sigma}^{D_1}| = |\mu_{\uparrow F}^{D_1}| |\vec{E}_{\sigma}|/\hbar$, where $\mu^{D_1}_{\uparrow,F}$ is the electric dipole moment:
\begin{equation}
\begin{split}
\mu^{D_1}_{\uparrow,F} &= e \bra{4^2S_{1/2},9/2,+9/2} \mathbf{r} \cdot \hat{\epsilon}_{\sigma} \ket{4^2P_{1/2},F,+9/2 + \sigma} \cr
\end{split}
\end{equation}
The $\hat{\epsilon}_\sigma$ represents the polarization direction of the eletric field. Because the strength of $E_{+1}^{(1)}$ is constant, the Raman field $M_{R1}$ can be represented as
\begin{equation}
M_{R1}(\mathbf{r}) = \mathrm{Re}[M_1( \mathbf{r})] \sigma_x + \mathrm{Im}[M_1(\mathbf{r})] \sigma_y,
\end{equation}
where $M_1(\mathbf{r})$ is proportional to $\sum_{j=1}^{3} \cos{ 2\mathbf{k}_{j} \cdot \mathbf{r} }$.

The blue Raman fields shown in Fig.\ref{fig:2}(b) are related to the strength $E_2$ and arise from the $4^2S_{1/2} \to 5^2P_{1/2}$ and $4^2S_{1/2} \to 5^2P_{3/2}$ transitions. Here, $E_{+1}^{(2)}(\mathbf{r}) = E_{R2}$ and $E_0^{(2)} = E_2 \sum_{j=1}^{3} \sin{ \mathbf{k}_{j} \cdot \mathbf{r} }$. The optical potential for spin $\tau$ can be calculated using the A.C. Stark effect and is denoted as
\begin{equation}
\begin{split}
V^{(2)}_{\tau}(\mathbf{r}) &= \sum_{F} \left( \frac{|\Omega^{5^2P_{3/2}}_{\tau F,+1}|^2 }{\Delta_{5^2P_{3/2}}} + \frac{|\Omega^{5^2P_{3/2}}_{\tau F,0}|^2}{\Delta_{5^2P_{3/2}}} \right) \cr
& + \sum_{F} \left( \frac{|\Omega^{5^2P_{1/2}}_{\tau F,+1}|^2}{\Delta_{5^2P_{1/2}}} + \frac{|\Omega^{5^2P_{1/2}}_{\tau F,0}|^2}{\Delta_{5^2P_{1/2}}} \right). \cr
\end{split}
\end{equation}
The Raman field caused in Fig.\ref{fig:2}(b) is
\begin{equation}
\begin{split}
M_{R2}(\mathbf{r}) &= \sum_F \left( \frac{\Omega^{5^2P_{3/2} \ast}_{\uparrow F,+1} \Omega^{5^2P_{3/2}}_{\uparrow F,0}}{\Delta_{5^2P_{3/2}}} + \frac{\Omega^{5^2P_{1/2} \ast}_{\uparrow F,+1} \Omega^{5^2P_{1/2}}_{\uparrow F,0}}{\Delta_{5^2P_{1/2}}} \right) \cr
&\left( \sum_{j=1}^{2} \delta( \Delta + \omega - \omega_{R2} - \delta_j(\mathbf{r}) ) \right) \ket{g_{\downarrow}}\bra{g_{\uparrow}} + h.c.,\cr
\end{split}
\end{equation}
The Raman field $M_{R2}$ can be represented as
\begin{equation}
M_{R2}(\mathbf{r}) = \mathrm{Re}[M_2(\mathbf{r})] \sigma_x + \mathrm{Im}[M_2(\mathbf{r})] \sigma_y,
\end{equation}
where $M_2(\mathbf{r})$ is proportional to $\sum_{j=1}^{3} \sin{ \mathbf{k}_{j} \cdot \mathbf{r} }$.

\section{Tight-binding parameters}
\label{app:4}

\begin{figure}[tbph]
    \centering\includegraphics[width=8cm]{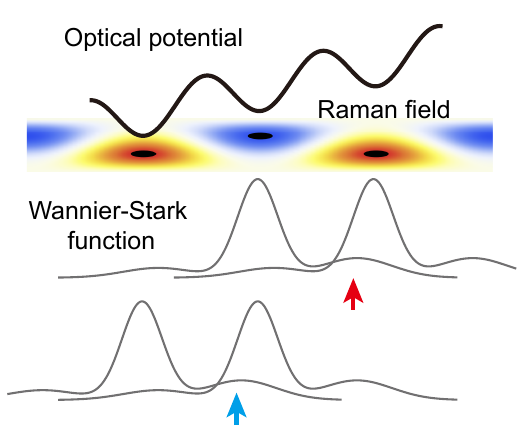}
    \caption{ Illustration of nearest-neighbor spin-flip terms caused by blue Raman fields. The imbalanced distribution of Wannier-Stark functions leads to staggered coupling.}
    \label{fig:5}
\end{figure}

The parameters of the tight-binding model can be calculated from the Wannier-Stark functions. We define the nearest-neighbor hopping with the Wannier-Stark state $w(\mathbf{r})$:
\begin{equation}
\begin{split}
t_{\mu} &= \int \mathrm{d}^2\mathbf{r}~ w_{B\sigma}^\ast(\mathbf{r} - \mathbf{R}_A - \mathbf{e}_{\mu}) \left( \frac{\mathbf{\mathbf{p}}^2}{2m} + V_{OL}(\mathbf{r}) \right) \cr
&\times w_{A\sigma}(\mathbf{r} - \mathbf{R}_A). \cr
\end{split}
\end{equation}
Mirror symmetry enforces $t_1 = t_2 \neq t_0$. The blue Raman fields $M_1(\mathbf{r})$ generates nearest-neighbor spin-flip terms
\begin{equation}
\begin{split}
g_{A\mu} &= \int \mathrm{d}^2\mathbf{r}~w_{B\downarrow}^\ast(\mathbf{r} - \mathbf{R}_A - \mathbf{e}_\mu) \left[ M_{R2,x}(\mathbf{r}) \right.\cr
& \left. + i M_{R2,y}(\mathbf{r}) \right] w_{A\uparrow}(\mathbf{r} - \mathbf{R}_A), \cr
g_{B\mu} &= \int \mathrm{d}^2\mathbf{r}~w_{A\downarrow}^\ast(\mathbf{r} - \mathbf{R}_B - \mathbf{e}_\mu) \left[ M_{R2,x}(\mathbf{r}) \right.\cr
& \left. + i M_{R2,y}(\mathbf{r}) + \right] w_{B\uparrow}(\mathbf{r} - \mathbf{R}_B),
\end{split}
\end{equation}
and for next-nearest-neighbor spin-flip hopping terms caused by red Raman fields $M_2(\mathbf{r})$:
\begin{equation}
\begin{split}
J_{A\mu\nu} &= \int \mathrm{d}^2\mathbf{r}~w_{A\downarrow}^\ast(\mathbf{r} - \mathbf{R}_A - \mathbf{e}_\mu + \mathbf{e}_\nu) \left[ M_{R1,x}(\mathbf{r}) \right. \\
& \left. + i M_{R1,y}(\mathbf{r}) \right] w_{A\uparrow}(\mathbf{r} - \mathbf{R}_A) \\
J_{B\mu\nu} &= \int \mathrm{d}^2\mathbf{r}~w_{B\downarrow}^\ast(\mathbf{r} - \mathbf{R}_B - \mathbf{e}_\mu + \mathbf{e}_\nu) \left[ M_{R1,x}(\mathbf{r}) \right. \\
& \left. + i M_{R1,y}(\mathbf{r}) \right] w_{B\uparrow}(\mathbf{r} - \mathbf{R}_B)
\end{split}
\end{equation}
where $J_{A\mu\nu} = -J_{B\mu\nu}$ since $C_6$ anti-symmetry of red Raman field $M_1(\mathbf{r})$.

Considering the actual case where the overlap center of two Wannier-Stark functions is not strictly located on the $C_6$ anti-symmetric line, we still need to account for the influence of these effects on the tight-binding model. This imply the red Raman fields also introduce nearest-neighbor spin-flip terms, which exhibit staggered coupling, as shown in Fig. \ref{fig:5}. The red and blue arrows indicate the main contribution regions of nearest-neighbor spin-flip hopping. Following the setting in the main text, by setting $J_1 = J_2 = i J_I$ and $g_1 = -g_2 = -i g$, the new terms cause by the properties of the Wannier-Stark functions can be considered as a modification of $g$, and this will not influence the form of the tight-binding model.

\nocite{*}
\bibliographystyle{apsrev4-1}
\bibliography{REV}

\end{document}